\documentclass[prb,aps,prl,twocolumn,showpacs,floatfix]{revtex4}

\usepackage{amsmath}
\usepackage{amssymb}
\usepackage{bm}
\usepackage{graphicx}

\def\be{\begin{equation}}
\def\ee{\end{equation}}
\def\bea{\begin{eqnarray}}
\def\eea{\end{eqnarray}}

\begin{document}

\title{
On the size scaling of the nearest level spacing at criticality 
}

\author{A.\ V.\ Malyshev}
\thanks{On leave from Ioffe Physiko-Technical Institute, 26 Politechnicheskaya str., 194021 Saint-Petersburg, Russia}
\affiliation{Departamento de F\'{\i}sica Aplicada, Universidad de Salamanca, E-37071 Salamanca, Spain}

\date{\today}

\begin{abstract}
It is conjectured that the size scaling of the nearest level spacing in the
critical spectral region, $S(N)\propto N^{-\lambda}$, remains qualitatively
the same within phases of extended and critical states. The exponent
$\lambda$ is therefore identical to that for the bare level spacing (at zero
disorder). Our calculation of the scaling for the one-dimensional model with
diagonal disorder and long-range power-like interaction confirms the
conjecture.
\end{abstract}


\maketitle

Localization-delocalization transition (LDT) in disordered systems,
predicted by Anderson for 3D in 1958~\cite{Anderson58}, still remains a
thoroughly investigated problem (see Refs.~\cite{Lee85,Kramer93} for an
overview). A remarkable progress has been achieved in understanding of
critical systems, in particular, the multifractal nature of critical wave
functions has been
discovered~\cite{Wegner80,Aoki83,Castellani86,Schreiber91,Janssen94}).
Statistics of energy levels within disordered models have also been
intensively investigated during the recent years. It was demonstrated that
statistical properties of level spacing are closely related to general
localization properties~\cite{Altshuler86,Shklovskii93,Kravtsov94}.

The spacing between two energy levels is determined by the overlap of the
corresponding wave functions. Thus, weak overlap of localized states results
in Poisson statistics of energy levels in the thermodynamic limit, which
implies that the level spacing can be arbitrary small and does not depend on
the system size. Contrary to that, strong overlap of the extended
eigenstates results in strong level repulsion. Therefore, Wigner-Dyson
statistics describe well energy levels in good
metal~\cite{Efetov83,Altshuler86}. The level spacing for extended states
depends on the system size. In this contribution we consider the level
spacing in a critical spectral regions, i.e. in a region where the
localization-delocalization transition takes place or at the mobility edge.

Within the phase of the extended states, the exponent $\lambda$ of the size
scaling of the level spacing, $S(N)\propto N^{-\lambda}$, is expected to be
independent of disorder, otherwise the statistics of the energy levels
spacing could change with disorder, which would be in direct contradiction
to the robustness of the Wigner-Dyson statistics within the phase of the
extended states~\cite{Efetov83,Altshuler86,Mirlin91}. At criticality, e.g.
close to the transition point or the mobility edge, the level statistics,
although manifesting strong level repulsion, is different from the
Wigner-Dyson one~\cite{Shklovskii93}. Moreover, reasonings based on the
consideration of wave function overlap should be used with caution, because
critical eigenstates are very sparse and it is not obvious how much they
overlap. Fyodorov and Mirlin addressed this problem within the framework of
the power law random-banded model (see Refs.~\onlinecite{Mirlin96,Mirlin00a}
and references therein for the detailed description of the model). They
showed analytically that two close critical eigenstates are strongly
correlated, in the sense that the overlap of the two states is of the order
of the self-overlap of a state~\cite{Fyodorov97}. It was conjectured in
Ref.~\cite{Fyodorov97} that the strong correlation of critical states is a
general property. This correlation results in strong level repulsion up to
the critical
point~\cite{Altshuler88,Shklovskii93,Evangelou94,Braun95,Zharekeshev95,Kravtsov94,Aronov95}.
Because the level spacing is determined by the overlap of wave functions, we
conjecture that the strong overlap of critical eigenfunctions preserves the
size scaling of the level spacing within the phase of the critical states
also. Thus, the scaling exponent $\lambda$ is expected to be the same
throughout both phases of extended and critical states. The latter phase
extends up to the magnitude of disorder at which the localization length
becomes equal to the linear system size. The exponent is therefore identical
to that for the bare level spacing (at zero disorder) in the considered
energy region.

To prove the conjecture on the independence of the scaling exponent of
disorder we consider the level spacing within the framework of the
tight-binding Hamiltonian on a one-dimensional (1D) regular lattice of $N$
sites with power-like inter-site interaction:
\begin{equation}
	{\cal H} = \sum_{n=1}^{N}\varepsilon_{n} |n\rangle\langle n| +
	\sum_{m,n=1}^{N}J_{mn}|m\rangle\langle n|\ ,
	\label{hamiltonian}
\end{equation}
where $|n\rangle$ is the ket vector of a state with on-site energy
$\varepsilon_{n}$. The energies are stochastic variables, uncorrelated for
different sites and distributed uniformly around zero within the interval of
width $W$. The hopping integrals are $J_{mn}=J/|m-n|^{\mu}$, $J_{nn}=0$ with
$1 < \mu \leq 3/2$. We set $J>0$, then the LDT occurs at the upper band edge
provided $1 < \mu < 3/2$~\cite{Rodriguez00,Rodriguez03}. The latter fact
makes the model very advantageous from the viewpoint of numerical
calculations because efficient algorithms, such as the Lanczos, can be used
to find a few extreme eigenstates of interest. In all calculations we
consider three uppermost states under open boundary conditions for system
sizes in the range $N=4096\div65536$. All results are averaged over more
than $5\, 10^3 \times (65536/N)$ disorder realizations. The lattice
constant is set to unity.

Recently, the model (\ref{hamiltonian}) was studied numerically for two
values of interaction exponent $\mu$: $\mu=3/2$ and
$\mu=4/3$~\cite{Malyshev03}. In the former marginal case no signs of
transition was found, while at $\mu=4/3$ the model was found to reveal the
LDT at $W_c = 10.9\pm0.2$. In this contribution we focus on the analysis of
the level spacing. In particular, we study the size scaling of the level
spacing at the top of the band in the vicinity of the critical value of
disorder $W_c$ for $\mu=4/3$.

We are particularly interested in the scaling of the level spacing within
the phase of the critical states. Because we deal with finite systems, the
phase of critical states is of finite width, so we determine the width
first. The critical phase extends from the critical point $W=W_c$ up to the
disorder magnitude $W=W_L$ at which the localization length becomes equal to
the linear system size: $\xi(W_L)=L$ (within the considered 1D model
$L\equiv N$). To determine the disorder magnitude $W_L$ we analyze the size
scaling of the typical value of the inverse participation ratio (IPR). The
IPR for the state $\alpha$ is defined in a standard way:
\begin{equation}
	{I_2^{(\alpha)}}=\sum_{n=1}^{N} |\psi_{\alpha n}|^{4}\ ,
	\label{pn}
\end{equation}
where $\psi_{\alpha n}$ is the $\alpha$-th normalized eigenstate of the
Hamiltonian~(\ref{hamiltonian}). The typical value of the IPR is defined as
$I_2=\exp\langle\,\ln{I_2^{(\alpha)}}\,\rangle$, where the angle brackets
stand for averaging over disorder realizations and $\alpha$. We use the
typical value of the IPR rather than the mean value of the IPR because,
unlike the latter, the former is a self-averaged
quantity~\cite{Evers00,Mirlin00b}.

Within the phase of extended states the IPR size scales as $I_2 \propto
L^{-1}$, while for the localized states the scaling reads: $I_2 \propto
L^{0}$. Wegner found that at criticality $I_2$ size scales
anomaluosly~\cite{Wegner80}: $I_2 \propto L^{-D_2}$ where the fractal
dimension $D_2$ is smaller than the spacial dimensionality, which results
from the multifractal structure of the
eigenfunctions~\cite{Wegner80,Aoki83,Castellani86,Schreiber91,Janssen94}.
Above the transition point the scaling of the IPR is governed by the
localization length $\xi$ rather than by the system size:
$I_2\propto\xi^{D_2}$~\cite{Fyodorov95}. The localization length depends on
the magnitude of disorder in the vicinity of the critical point
$\xi(W)=\xi_0|W/W_c-1|^{-\nu}$ ($\nu>0$), therefore, two scaling regimes are
expected above the critical point. Within the phase of the critical states,
i.e. for $W>W_c$ and system sizes $L<\xi(W)$, the IPR scaling is still
determined by the size of the system: $I_2 \propto L^{-D_2}$. Contrary to
that, within the phase of the localized states, i.e. for $L>\xi(W)$, the IPR
scaling is determined by the localization length $\xi(W)$ and does not
depend on the system size.

\begin{figure}[ht!]
\includegraphics[width=\columnwidth,clip]{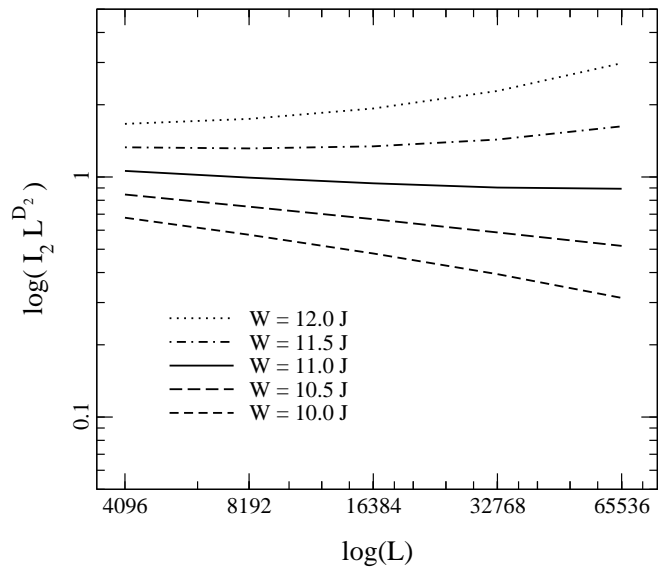} 
\caption{
	Size scaling of the parameter $\widetilde{I}_2=I_2\,L^{D_2}$
	in the vicinity of the critical point. 
	}
	\label{I2xND2_SizeScaling}
\end{figure}

Fig.~\ref{I2xND2_SizeScaling} shows the size scaling of the quantity 
$\widetilde{I}_2=I_2\,L^{D_2}$. The value of the correlation dimension $D_2$ was
obtained by studying the size scaling of the typical value of the IPR at
criticality: $D_2=0.66$ for $\mu=4/3$ (a detailed analysis of the
correlation dimension is to be published elsewhere). Within the phase of the
extended states the quantity $\widetilde{I}_2$ decreases with the system
size: $\widetilde{I}_2\propto L^{D_2-1}$ ($D_2<1$). At the transition point,
$\widetilde{I}_2$ is expected to be constant. Above the transition, two scaling
regimes are expected: $\widetilde{I}_2$ remains constant for $L<\xi(W)$, while
for systems that are larger than the localization length, $\widetilde{I}_2$ is
expected to increase with the size: $\widetilde{I}_2\propto\xi^{-D_2}\,L$. The
above transparent scenario suggests that for a given set of system sizes the
phase of the critical states extends from the critical point $W=W_c$ up to
the disorder magnitude $W=\widetilde{W}>W_c$ at which the size scaling of
$\widetilde{I}_2$ starts to deviate from a constant (as a matter of fact,
$\widetilde{W}=W_{L}$ for $L=L_{max}$). Fig.~\ref{I2xND2_SizeScaling}
demonstrates that, for the considered system sizes ($L\equiv
N=4096\div65536$), the critical phase extends up to the $\widetilde{W}
\approx 11.5\,J$.

Having obtained the disorder at which the uppermost eigenstates are critical
we turn to the analyses of the size scaling of the level spacing for these
states. Under periodic boundary conditions the level spacing at the upper
edge of the {\it bare} band (for non-disordered system) depends on the
system size $N$ ($N \gg 1$) as follows:
\begin{equation}
	S_p(N) = \frac{C(\mu)}{N^{\mu-1}}
	+ O\left(\frac{1}{N^{\mu}}\right)
	\ ,
	\label{SconN}
\end{equation}
where $C(\mu)=2\,\Gamma(2-\mu)\,\cos(\pi\,(\mu-1)/2)$.
Under open boundary conditions, which we used in all calculations, no
analytical expression for the spacing at the upper band edge can be obtained.
Nevertheless, its size scaling is close to Eq.~(\ref{SconN}) and the leading
(non-zero) power of $N$ in the expansion is also $1-\mu$. In particular, for
$\mu = 4/3$ the spacing reads:
\begin{equation}
	S(N) = \frac{C_{4/3}}{N^{1/3}}+ O\left(\frac{1}{N^{4/3}}\right)\ ,
	\label{SonN}
\end{equation}
the constant $C_{4/3}$ was obtained numerically: $C_{4/3}\approx 3.58$

\begin{figure}[ht!]
\includegraphics[width=\columnwidth,clip]{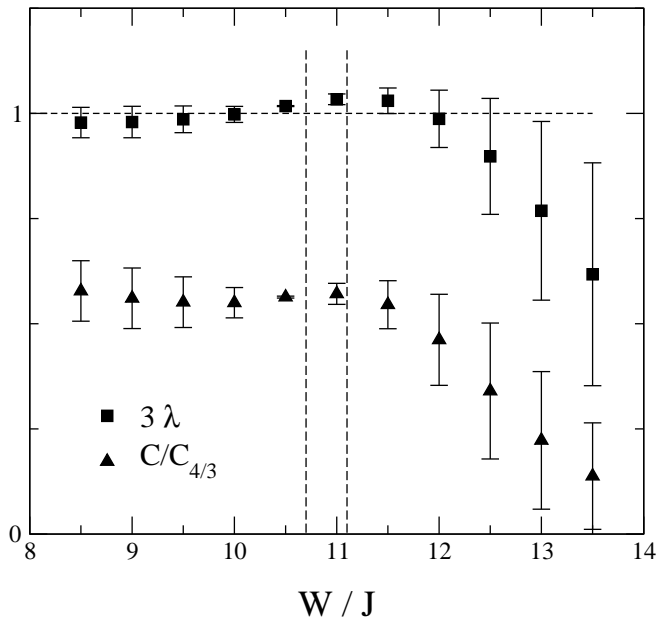} 
\caption{
	Disorder scaling of the parameters $\lambda$ and $C$ for $\mu=4/3$
	in the vicinity of the critical point. 
	}
	\label{LS_SizeScaling}
\end{figure}

We calculated the size scaling of the mean level spacing at the upper band
edge for $\mu=4/3$ and different values of disorder magnitude $W$. For each
value of $W$, we fit the formula $C\,N^{-\lambda}$ to the calculated size
dependence of the mean spacing, varying the system size within the range
$N=4096\div65536$. The result of the fit is presented in
Fig.~\ref{LS_SizeScaling}: squares show the disorder dependence of the
exponent $\lambda$ normalized to its value for non-disordered system, $1/3$,
while triangles show the disorder scaling of the factor $C$ normalized to
$C_{4/3}$ (see Eq.~\ref{SonN}). Vertical lines demonstrate the critical
disorder $W=10.7\div 11.1\,J$~\cite{Malyshev03}. Error bars represent
confidence intervals of the parameters $C$ and $\lambda$. The statistical
error is of the order of the symbol size.

The figure demonstrates that within the phase of the extended and critical
states, that is for $W<11.5\,J$, the exponent that determines the size
dependence of the mean level spacing in the critical energy region (i.e. at
the upper band edge) is the same as that for the bare band: $\lambda=1/3$.
The disorder reduces the level spacing, as expected, but does not affect the
qualitative dependence of the spacing on the system size. Within the phase
of the localized states, that is for $W>11.5\,J$, the localization length
becomes smaller than the system size, and the level spacing is {\it not}
determined by the size of the system. Therefore, the formula
$C\,N^{-\lambda}$ becomes irrelevant within the phase of the localized
states; the increase of the confidence intervals for the parameters $C$ and
$\lambda$ in the range $W>11.5\,J$ manifests the irrelevance.

To conclude, we have demonstrated that within both phases of the extended
and critical states the size scaling of the level spacing remains
qualitatively unchanged: the exponent of the size scaling does not depend on
disorder. The exponent is therefore the same as that in the considered
energy region of the bare band (for a non-disordered system). The above
result is probably general as it is a consequence of strong overlap of
critical eigenstates that was reported on recently~\cite{Fyodorov97}. The
independence of the scaling exponent of the disorder has several important
consequences. It suggests, in particular, that the bare scaling of the
energy levels can be used for qualitative argumentations that are applied to
disordered systems (see, for example, Ref.~\onlinecite{Rodriguez03} and
references therein). Properties of critical states are usually studied by
considering all states within an energy window in the vicinity of the
critical point. The mean level spacing $\Delta\propto N^{-1}$ is often used
to estimate the number of critical states, which can lead to an
over-estimation of this number, because the spacing in the critical region
is much greater than $\Delta$. In particular, for the standard
three-dimensional Anderson model the critical spectral region is the center
of the band, where the spacing scales as $S\propto N^{-1/3}$; the relevant
spacing is therefore much greater than the average one. The over-estimation
of the number of the critical states can lead to consideration of not
only critical states but localized states also, which would result in the
under-estimation of the correlation dimension, affect the energy level
statistics, etc.

\begin{acknowledgments}
The author thank V.\ A.\ Malyshev and F.\ Dom\'{\i}nguez-Ada\-me for
discussions. This work was supported by MECyD (SB2001-0146).
\end{acknowledgments}

\end{document}